\documentclass{jfm}
\usepackage{graphicx}
\usepackage{epstopdf, epsfig}

\newcommand{\pt}{\partial}
\newcommand{\cs}{c}
\newcommand{\cso}{c_0}
\newcommand{\calR}{{\cal R}}
\newcommand{\drop}{d_{r_0}^+} 
\newcommand{\drom}{d_{r_0}^-} 
\newcommand{\dropm}{d_{r_0}^\pm} 
\newcommand{\bfa}{}

\usepackage{color}

\newcommand{\cdmnew}{}

\shorttitle{1D acoustics in inhomogeneous atmospheres}
\shortauthor{C. D. Matzner and S. Ro}

\title{On Linear and Nonlinear Acoustics in Stratified, Variable-Area Ducts and Atmospheres, and Lighthill's Proposition}

\author{C.\ D.\ Matzner\aff{1}
  \corresp{\email{matzner@astro.utoronto.ca}},
 \and S. Ro\aff{2}}

\affiliation{\aff{1}David A.\ Dunlap Department of Astronomy and Astrophysics, University of Toronto, 50 St.\ George Street, Toronto M6J 2K2, Canada
\aff{2}Astronomy Department and Theoretical Astrophysics Center, University of California, Berkeley, Berkeley, CA 94720, USA}

\begin{document}

\maketitle

\begin{abstract}
We consider linear and nonlinear waves in a stratified hydrostatic fluid within a channel of variable area, under the restriction of one-dimensional flow.  We derive a modified version of Riemann's invariant that is related to the wave luminosity.  This quantity obeys a simple dynamical equation in linear theory, from which the rules of wave reflection are easily discerned; and it is adiabatically conserved in the high-frequency limit.  Following a suggestion by Lighthill, we apply the linear adiabatic invariant to predict mildly nonlinear waves.  This incurs only moderate error.  We find that Lighthill's criterion for shock formation is essentially exact for leading shocks, and for shocks within high-frequency waves.  We conclude  that approximate invariants can be used to accurately predict the self-distortion of low-amplitude acoustic pulses, as well as the dissipation patterns for weak shocks, in complicated environments such as stellar envelopes.   We also identify fully nonlinear solutions for a restricted class of problems. 
\end{abstract}

\section{Introduction}\label{S:intro} 

The propagation of finite-amplitude acoustic waves, and especially the creation of shocks within them, are of interest for physical problems across a wide range of scales -- from the collapse of sonoluminescent bubbles \citep{2001JFM...431..161L} to the eruption of stellar envelopes in pre-supernova outbursts \citep{smith2013model}.
We shall consider the case of quasi-one-dimensional oscillations of an otherwise hydrostatic, stratified background state, allowing for variations in the cross-sectional area of the channel or atmosphere.

An important point of reference is the simplest limit in which the cross-sectional area is constant and the background state is uniform. \citet{riemann1860ShockFormn} solved this case exactly, up to the moment of shock formation within the flow, in terms of invariants conserved along sound fronts (characteristic trajectories).  For the subsequent evolution, \citeauthor{whitham1975linear} (\citeyear{whitham1975linear}, building on the work of  \citealt{landau1945shock}) developed an elegant analytical formalism to very accurately predict the strength and trajectory of a weak shock. Specifically, Whitham's `equal area' construction applies to shocks made by a disturbance traveling in one direction only, i.e., a simple wave.  

Can these solutions be adapted to the general problem?  In chapter 2 of {\em Waves in Fluids} \citep{lighthill2001WavesInFluids}, M.~J.~Lighthill proposed that they can. Considering the equivalent of simple waves, Lighthill proposed an invariant invariant should be approximately conserved along characteristic trajectories -- approximately, because the argument relies on low-amplitude perturbations and on the high-frequency limit in which wave reflection is negligible.   Coupled with the  pressure-velocity relation appropriate to simple waves, this conservation law allows one to transform a traveling-wave problem in the general problem to its analog in the simple limit, for which Riemann's and Whitham's approaches provide a solution.

This method is appealing as an analytical technique even though it is not exact.  Low-amplitude, high-frequency waves are precisely the type that sample variations in their environment before forming a shock. Furthermore, such waves are challenging to treat with direct numerical simulations because accuracy demands hundreds or thousands of numerical zones per wavelength.  A shock forms only after many wavelengths of propagation if the wave amplitude is low. 

While Lighthill's proposal has not, to our knowledge, been re-examined, similar approaches have certainly been considered.  High-frequency scalings, familiar from WKB theory, motivate \citet{subrahmanyam2001family}'s proposed exact solutions for specific problems in linear acoustics. In our own previous work \citep{2017ApJ...841....9R}, we reinvented aspects of Lighthill's proposal, including his criterion for shock formation (his eq.~254), by positing that the high-frequency scalings apply to individual characteristics.   Other authors, such as \citet{2001JFM...431..161L}, \citet{tyagi2003nonlinear}, and \citet{tyagi2005propagation}, also derive shock formation criteria that coincide with Lighthill's, as \citeauthor{tyagi2003nonlinear} note.

To develop the theory further, we shall express linear acoustical motions in terms of wave amplitudes that are conserved along characteristics except for the effects of reflection.  These amplitudes are modified versions of Riemann's invariants; they are equivalent to Lighthill's proposed invariant in a linear travelling wave; and they are related in a simple way to the wave luminosity.  When the linearized amplitude equation is applied to a nonlinear wave, errors are suppressed by factors of the gradients in the background.  This reflects the fact that wave amplitudes reduce to Riemann's invariants in the uniform case.  Together, these points prove the validity of Lighthill's proposal and the exactness of his shock formation criterion in certain contexts, while also providing a convenient way to evaluate wave reflection. 

We start with adiabatic, quasi-one-dimensional fluid equations in Lagrangian form, and obtain Riemann's invariants $J_\pm$ for the simple case of planar motion of a uniform fluid.  We then derive the variation of $J_\pm$ in the general problem.  Linearizing this result and introducing wave amplitudes $\calR_\pm$, we arrive at the linear amplitude evolution equation.  We relate these amplitudes to the wave luminosity, and consider the dynamics of wave reflection.   We then examine the nonlinear errors incurred when the linearized equation is applied to waves of finite amplitude, showing that these are small in the high-frequency limit.  Moreover, we show that Lighthill's shock formation condition is exact in certain cases, and essentially exact for shocks forming within high-frequency waves.  Then, we demonstrate that nonlinear solutions are available for a certain class of problems.   Finally, we conduct numerical experiments to demonstrate the application to traveling waves in spherical and strongly stratified environments.

\subsection{\cdmnew Notation} Our coordinates are radius $r$ and time $t$.  The gravitational acceleration $g(r)$, the cross-sectional area $A(r)$, the ratio of specific heats $\gamma(r_0)$, and the background sound speed $\cso(r_0)$ are all considered arbitrary continuous functions.   Fluid variables include pressure $P$, density $\rho$, velocity $v$, and sound speed $\cs$. From these, we construct Riemann invariants $J_\pm$ and wave amplitudes $\calR_\pm$, as well as the right- and left-going wave luminosities $L_{w\pm}$ along characteristic trajectories $C^\pm$. The net wave luminosity is ${\cal L}_w = L_{w+}-L_{w-}$.  An unperturbed quantity in the hydrostatic background state has subscript `0', and Lagrangian perturbations (for the same fluid element) have prefix `$\delta$', i.e., $\delta f = f - f_0$.  
For convenience, we define $x\equiv \rho/\rho_0$, and $q\equiv(\gamma-1)/2$.  We also define $\varepsilon$ as the relative strength of a reflected wave and $\tau$ as the time of propagation in \S~\ref{SS:Reflect-highfreq}.  The characteristic acoustic impedance is $Z$.  

We use operator and subscript notation for partial derivatives, so $\partial_t f$ and $f_{t}$ both mean $\partial f/\partial t$.  
The Lagrangian time derivative operator is $d_t$, equivalent to $\partial_t + v\partial_r$ when we use variables $(r,t)$, and equivalent to $\partial_t$ when our variables are $(r_0,t)$.   Subscripts $+$ and $-$ indicate that a quantity is associated with $C^+$ or $C^-$, i.e., sound fronts moving to increasing or decreasing $r$. A dot indicates the time derivative along the appropriate trajectory: so $\dot J_+ = [\partial_t + (v+\cs)\partial_r] J_+$.   Radial derivatives along $C^\pm$ trajectories are denoted $\dropm$. 

Note, we do not address the phenomena that excite waves in the hydrostatic background state.  The acoustical perturbations we study must therefore enter the zone of interest across its boundary, or be the result of body forces that upset the initial equilibrium but have since disappeared. 

\section{Equations of motion} \label{S:equations} 

Our quasi-one-dimensional equations incorporate a couple assumptions: first, that motions transverse to the gradient of $r$ can be ignored; and second, that displacements are small enough ($\delta r\ll r_0$) that smooth functions of $r$ can be replaced with their values at $r_0$:  for instance, $g=g_0$ and $A=A_0$.
In addition, we neglect dissipative effects such as thermal diffusion; see \citet{2017ApJ...841....9R} for a justification in the context of stellar outbursts. 

We start with the conservation of mass,
\[ d_t \ln \rho + A^{-1} \partial_r A v =0 , \] 
 momentum,
\[ d_t v + \frac1\rho \partial_r P = -g,\]
as well as background hydrostatic equilibrium,
\[ \frac1\rho_0 \partial_{r_0} P_0 = -g_0.\] 
We consider the fluid to be adiabatic and for simplicity we adopt an ideal gas equation of state, 
\[ P/P_0 = (\rho/\rho_0)^\gamma = x^\gamma, \]
where $\gamma$ may be a function of $r_0$.

We convert the spatial gradient ($\partial_r$) into gradients with respect to the initial radius ($\partial_{r_0}$) using $A \rho\, dr = A_0 \rho_0\, dr_0$, which implies 
\[\rho^{-1} \partial_r = (A/A_0) \rho_0^{-1} \partial_{r_0} \simeq \rho_0^{-1} \partial_{r_0} \] 
 using $A\simeq A_0$.  At the same time, we switch spatial variables from $r$ to $r_0$, allowing $d_t \rightarrow \partial_t$.  Our equation for mass conservation becomes (using $r^{-1} \simeq r_0^{-1}$), 
\[ \partial_t \ln (\rho/\rho_0) + (\rho/\rho_0) \partial_{r_0} v + v \,\partial_{r_0} \ln A = 0.\] 
The momentum equation becomes 
\[ \partial_t v + \frac1{\rho_0} \partial_{r_0} \delta P = 0 \]
or, using  $P = x^\gamma P_0$ and   
writing out the derivative, 
\[ \partial_t v +  \frac{\gamma P_0}{\rho_0} x^{\gamma-1} \partial_{r_0} x +  \frac{\partial_{r_0} P_0}{\rho_0} (x^\gamma-1) + \cso^2 (\ln x)\partial_{r_0}\ln \gamma  = 0, \] 
which we  re-write, using $\gamma P_0= \rho_0 c_0^2$, $\partial_{r_0} P_0 =- \rho_0 g_0$, 
and $c_0^2 x^{\gamma-1} = \cs^2$, as 
\[ \partial_t v + \cs^2  \partial_{r_0} x =  (x^\gamma-1)  g_0 - \cso^2(\ln x)\partial_{r_0}\ln \gamma.  \]
The mass equation takes in the form 
\[\partial_t  x + x^2 \partial_{r_0} v  = - v x\, \partial_{r_0} \ln A,
 \]
where we have multiplied through by $x$ to put the time derivatives in the same form as the momentum equation.   Combining the momentum and mass equations, 
\begin{equation}\label{eq-matrix} \partial_t \mathbf{u} + \mathsfbi{B}\, \partial_{r_0} \mathbf{u} = \mathbf{s}, \end{equation}
 where 
 \[ \mathbf{u} = \left(\begin{array}{l}
   v \\ x\end{array} \right),
   ~~\mathsfbi{B} = \left[ \begin{array}{ll}  0 &\cs^2 \\ x^2 & 0 \end{array} \right], 
~~{\rm and}~~ \mathbf{s} = \left(\begin{array}{c} (x^\gamma-1) g_0- \cso^2(\ln x)\partial_{r_0}\ln \gamma \\
- v x\, \partial_{r_0} \ln A\end{array}\right). \] 

Note that our approximation of small radial perturbation only affects the form of $\mathbf{s}$. 

\section{Uniform planar case: Riemann invariants} \label{S:Planar-Riemann}

In the simplest limit of a homogeneous background fluid and constant area, $g_0=0$ and $\partial_r \ln A=0$ so that $\mathbf{s}=0$.  To derive the Riemann invariants, we seek a function $J(x,v)$ and a Lagrangian speed $\lambda$ that satisfy $\partial_t J + \lambda \partial_{r_0} J =0$.   

Writing out  these partial derivatives in subscript notation, $ J_t = J_x x_t + J_v v_t = (J_v, J_x) \mathbf{u}_t$ and likewise $J_{r_0} = (J_v, J_x)\mathbf{u}_{r_0}$.   The equation we wish to solve, $J_t + \lambda J_{r_0}=0$, becomes $(J_v, J_x)\mathbf{u}_t + \lambda (J_v, J_x)\mathbf{u}_{r_0} =0$; using (\ref{eq-matrix}) to  eliminate the time derivative,
\[ (J_v, J_x) ( \mathsfbi{B}  - \lambda \mathsfbi{I} ) \mathbf{u}_{r_0} =0.\]   
For this to be true for any $\mathbf{u}_{r_0}$ requires $(J_v, J_x) ( \mathsfbi{B}  - \lambda \mathsfbi{I} )=0$. Taking the transpose, 
\[ ( \mathsfbi{B} ^T - \lambda \mathsfbi{I})(J_v, J_x) ^T=0, \] 
which is solved if $\lambda$ and $(J_v, J_x)^T$ are an eigenvalue and corresponding eigenvector of $\mathsfbi{B} ^T$, respectively.  

The eigenvalues are $\lambda_\pm = \pm \cs x$.  These represent the Lagrangian speeds of sound fronts moving to the right or left though space at speed $v\pm \cs$, because $A_0 \rho_0 (d r_0/dt) =A \rho (d r/dt - v)$ along any trajectory, and  $A\rightarrow A_0$ for small perturbations.   (We refer to $\lambda_\pm$ variously as the wave speed, propagation speed, or characteristic speed, and to trajectories $C^\pm$ obeying $\dot r_0 = \lambda_\pm$ as characteristics or sound fronts.)

The corresponding eigenvectors are $(J_v, J_x)_\pm = (1, \pm \cs/x)$.  The functions that have these partial derivatives are the usual Riemann quantities,
\[J_\pm = v \pm \int \cs \frac{dx}{x}  = v \pm \frac{\cs}q,\]
which are invariants when $\mathbf{s}=0$. Having reconstructed the conservation of the Riemann quantities for this restricted problem, we now turn to how $J_\pm$ vary in the more general context. 

\section{General problem}\label{S:general} 
Lifting the restrictions of planar flow and a uniform fluid, Riemann's quantity $J$ is no longer invariant. Its time variation along a characteristic is 
\[\dot J = {\sum_{f\in \{v,x,\cso,q\} }}J_f(f_t + \lambda f_{r_0}) = (J_v,J_x) \cdot \mathbf{s} + \lambda J_{\cso} \partial_{r_0} \cso + \lambda J_q\partial_{r_0} q.  \] 
The first term on the right hand side arises from non-zero gravity and changes of area (components of $\mathbf{s}$), while the second and third terms account for gradients of the fluid properties; these contain only radial derivatives because $\cso$ and $\gamma$ are constant within each fluid element. 

We prefer to work with radial derivatives for later convenience, which we compute as $d_{r_0} J = \dot J/\lambda$.   Written out for outward and inward waves, 
\[ \dropm{J_\pm} = \pm\frac{x^\gamma-1}{x^{(\gamma+1)/2}} \frac{g_0}{\cso} 
\mp \frac{\cso^2 \ln x}{\cs x}\partial_{r_0}\ln\gamma
- \frac{v}{x}\partial_{r_0} \ln A
\pm \frac{\cs}{q}\partial_{r_0} \ln \cso
\mp \frac{\cs}{q} (1-q\ln x) \partial_{r_0} \ln q.\]
Part of the change in $J_\pm$ is acoustical, but part is due to the radial gradient of the unperturbed state.  To isolate the acoustical portion, we subtract $\partial_{r_0} J_{0\pm}$ to obtain
\begin{equation}\label{eq:dJdr_Unlinearized}
\dropm \delta J_\pm = 
\pm\frac{x^\gamma-1}{x^{(\gamma+1)/2}} \frac{g_0}{\cso} 
\mp \frac{\cso^2 \ln x}{\cs x}\partial_{r_0}\ln\gamma
- \frac{v}{x}\partial_{r_0} \ln A
\pm \frac{\delta\cs}{q}\partial_{r_0} \ln \cso
\mp \left(\frac{\delta\cs}{q}- \cs\ln x\right)  \partial_{r_0} \ln q.
\end{equation} 
Equation (\ref{eq:dJdr_Unlinearized}) is exact within the quasi-one-dimensional approximation we have adopted.  It therefore provides a point of comparison for the linearized wave amplitude equation derived below.

\section{Linear acoustics: wave amplitude and luminosity} \label{SS:calRandLw}
At this point we focus on linear perturbations and expand to leading order in $\delta x = x-1$: 
\[
\dropm \delta J_\pm = 
\pm\gamma\, \delta x \frac{g_0}{\cso} 
\mp \cso \delta x \,\partial_{r_0} \ln \gamma
- v\partial_{r_0} \ln A
\pm \frac{\delta\cs}{q}\partial_{r_0} \ln \cso
\mp \left(\frac{\delta\cs}{q}- \cso\,\delta x\right)  \partial_{r_0} \ln q.
\] 
The first term, due to gravity, can be rewritten using $\gamma g_0/\cso = - \cso \partial_{r_0} \ln P_0$.   

We now express the perturbations in terms of $\delta J_\pm$ using $v = (\delta J_+ + \delta J_-)/2$ and $(\delta c)/q =  (\delta J_+ - \delta J_-)/2 = c_0 \delta x + {\cal O}(\delta x^2)$. The term involving $\partial_{r_0}\ln q$ is zero to leading order, and the rest can be written compactly, using $\gamma P_0 = \rho_0\cso^2$ and $\delta J_+-\delta J_- = \pm(\delta J_\pm-\delta J_\mp)$, as  
\begin{equation}\label{eq:acoustics-dJ} 
\dropm \delta J_\pm = 
\delta J_\pm\, \partial_{r_0} \ln \frac{1}{\sqrt{\rho_0 \cso A}}+ \delta J_\mp \,\partial_{r_0} \ln  \sqrt{Z}
\end{equation}
{\bfa where $Z = \rho_0 c_0/A$ is the characteristic acoustic impedance. }
The first term represents the inward or outward propagation of an acoustic disturbance, while the second term encapsulates a conversion between it and the counter-propagating disturbance in the process of reflection, which is catalyzed by gradients of $Z$.

Equation (\ref{eq:acoustics-dJ}) is simplified in terms of a new variable, the wave amplitude: 
\begin{equation} \label{eq:defR}
\calR_\pm \equiv \sqrt{A\rho_0 \cso}~ \delta J_\pm.
\end{equation}
In linear theory this quantity evolves along $C^\pm$ only due to reflection: 
\begin{equation} \label{eq:acoustics-calR} 
\dropm \calR_\pm = 
 \calR_\mp \,\partial_{r_0} \ln  \sqrt{Z},
\end{equation}
so it is effectively conserved so long as reflection is negligible.  This  wave amplitude equation provides a means to solve linear acoustics in the general problem by the method of characteristics. As we shall see, it also provides a connection to Lighthill's approximate invariant and to the wave luminosity.  Moreover, it provide a useful means to approximate nonlinear acoustics as well. 

\subsection{Quasi-simple waves: equipartition and Lighthill's invariant} \label{SS:QuasiSimple}

A `simple wave' in the planar homogeneous problem is one in which one wave family has constant amplitude, so that every physical quantity is determined by the other's amplitude.   The analogous situation in the general problem is the case $|\varepsilon|\ll 1$, where $\varepsilon \equiv {\calR_\mp}/{\calR_\pm} = {\delta J_\mp}/{\delta J_\pm}$  measures the relative amplitude of the counter-propagating wave. We call this a `quasi-simple' wave.  From the definition $J_\pm$, 
\[ v = \pm \frac{1-\varepsilon}{1+\varepsilon}\frac{\delta \cs}q\rightarrow  \pm\frac{\delta \cs}q, \] where the limit holds for $|\varepsilon|\rightarrow 0$.  To linear order, quasi-simple waves obey the condition $\delta P = \pm \rho_0 c_0 v$ for equipartition between  kinetic and thermal energy perturbations, as well as the relation $v/\cso = \pm \delta x$. 

Motions can be decomposed into quasi-simple waves whenever wave reflection is negligible. By equation (\ref{eq:acoustics-calR}), this includes any environment in which $Z$ is constant or varies only over many wavelengths; see \S~\ref{SS:Reflect-highfreq}. 


\citet{lighthill2001WavesInFluids} identified $\delta P/\sqrt{Z}$ as an adiabatic invariant in the high-frequency limit.   Lighthill's invariant is equivalent to our $\pm \calR_\pm/2$, to linear order, within a quasi-simple wave in which $R^\mp=0$. 


\subsection{Wave luminosity}\label{SS:Lw}
We may relate $\calR_\pm$ to a linear estimate for the  energy flow carried by the wave, i.e., the instantaneous wave luminosity: 
\begin{equation}\label{eq:def-Lw} L_{w\pm} =\frac14 \calR_\pm^2
= \frac14 A \rho_0 \cso (\delta J_\pm)^2.  \end{equation} 
We define $L_{w\pm}$ to be positive, regardless of the direction of propagation.  
The coefficient $1/4$ arises because equipartition implies a total wave energy $v^2$ per unit mass, whereas $(\delta J_\pm)^2 = 4 v^2$ when $v = \pm \delta c/q$. 

The evolution of wave luminosity along a characteristic follows directly from equation (\ref{eq:acoustics-calR}):
\begin{equation}\label{eq:acoustics-Lw}  
\dropm
\ln L_{w\pm}= \varepsilon \,\partial_{r_0} \ln  Z.
\end{equation}



The net acoustic luminosity ${\cal L}_w$ is the difference between outward and inward wave luminosities: 
\begin{equation} \label{eq:def-net-Lw}
{\cal L}_w = L_{w+} - L_{w-} =  A \rho_0 \cso \frac{\delta \cs}{q} v \simeq A v\,\delta P,
\end{equation}
where the last approximation, which holds to linear order, follows from $P/P_0 = (\cs/\cso)^{\gamma/q}$ and $\gamma P = \rho\cs^2$.   Equation (\ref{eq:def-net-Lw}) specifies ${\cal L}_w$ as  the rate of work done against the pressure perturbation.  



\subsection{Wave reflection: low-frequency limit} \label{S:Reflection}

\begin{figure}
  \centerline{
  \includegraphics[width=\textwidth]{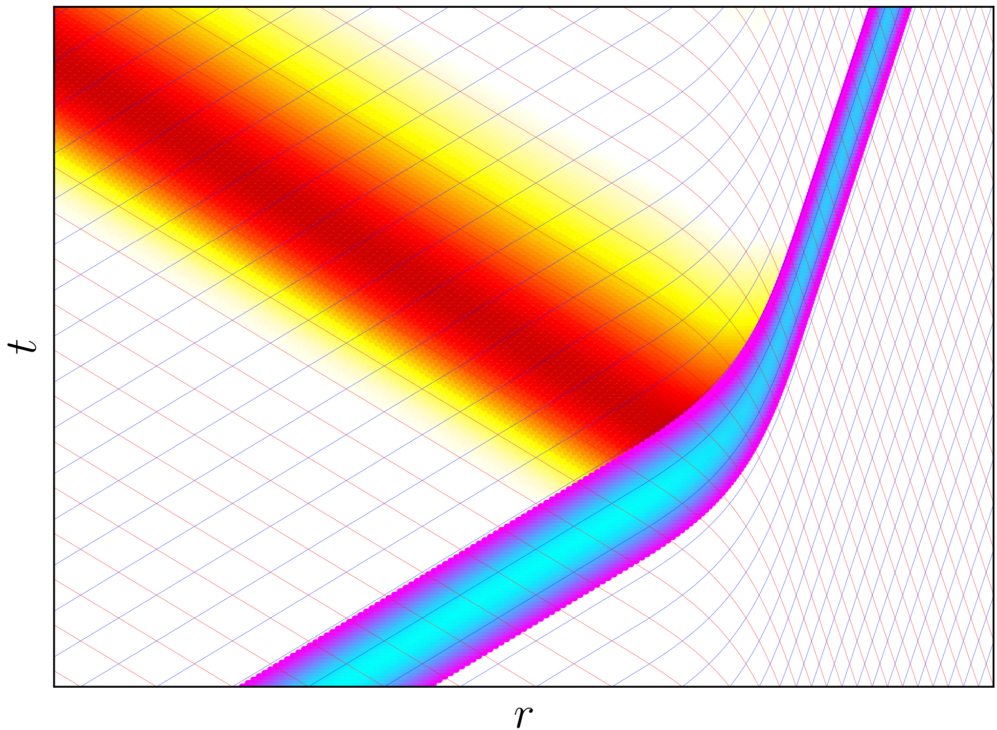}
  }
\caption{\cdmnew A schematic of wave reflection computed via the method of characteristics using equation (\ref{eq:acoustics-calR}).  Right and left-going characteristics (thin blue and red lines) cross a zone of varying sound speed and density within which the impedance varies, while $P$ and $A$ are constant.  A right-going acoustic pulse (purple and blue zone, depicting positive $\calR_+$) encounters the region of varying impedance, generating a reflected pulse (yellow and red zone, depicting negative values of $\calR_-$). The pulse is assumed to be weak enough that the perturbation to the wave speed is negligible. }
\label{fig:reflection}
\end{figure}

Equation (\ref{eq:acoustics-calR}) describes reflection, a process we depict in Figure \ref{fig:reflection}. It must, therefore, recover the known high and low-frequency limits of this process. 


In the low-frequency limit, we use the fact that $\dropm = (x\cs)^{-1}\partial_t \pm  \partial_{r_0}$, but $(x\cs)^{-1} \partial_t$ is negligible relative to $\partial_{r_0}$ in this limit.  Equation (\ref{eq:acoustics-calR})  becomes 
\begin{equation}\label{eq:LowFreqReflection} 
 \partial_{r_0} \calR_\pm = \calR_\mp \, \partial_{r_0} \ln \sqrt{Z},
 \end{equation}
for which the solution is $\calR_\pm = a\sqrt{Z} \pm b/\sqrt{Z}$ for constants $a$ and $b$. 
Consider an outward-propagating disturbance that encounters a zone separating a uniform inner region 1 (impedance $Z_1$) from a uniform outer region 2 (impedance $Z_2$).  The incident and reflected wave amplitudes are $\calR_{+,1}$ and $\calR_{-,1}$, respectively; the transmitted amplitude is $\calR_{+,2}$; but there is no inward wave in the outer region: $\calR_{-,2}=a\sqrt{Z_2}-b/\sqrt{Z_2} = 0$, so $a/b = Z_2$. One then finds that 
\[ \frac{ \calR_{+,1} }{ \calR_{-,1} } = \frac{Z_1 + Z_2}{Z_1 - Z_2}.  \]
The ratio of incident to reflected wave luminosities is therefore 
\[ \frac{ L_{w+,1} }{L_{w-,1} } = \frac{(Z_1 + Z_2)^2}{(Z_1 - Z_2)^2}, \] 
which is the classical result for reflection at a discontinuity in $Z$ \citep[e.g.,][]{campos1986waves}.

\subsection{Wave reflection: high-frequency limit} \label{SS:Reflect-highfreq}

We consider again the problem of a wave emanating outward from small $r$ and encountering a change in $Z$. Knowing in advance that reflection is minimal at high frequencies for which the wavelength of oscillation is much shorter than the scale of changes in $Z$, we apply an iterative procedure in which reflection is neglected at zeroth order ($\drop \calR_+^{(0)}=0$; $\calR_-^{(0)}=0$) and subsequent iterations obey $\drop \calR_\pm^{(i)} =\calR_\mp^{(i-1)} \partial_{r_0} \ln\sqrt{Z} $. 

We label outward and inward characteristics by the time $t_{i\pm}$ at which they cross a fiducial radius $r_{0i}$.  Along the outward ones, $t(r_0) = t_{i+} + \tau$, where $\tau = \int_{r_{0i}}^{r_0} dr_0'/(x\cs)$ is the propagation time; and along inward ones, $t(r_0) = t_{i-} - \tau$.  Focusing on linear amplitudes, for which the perturbation to the wave speed can be ignored, we take $\tau\rightarrow\tau_0(r_0) = \int_{r_{0i}}^{r_0} dr_0'/\cso(r_0')$. 

For the problem at hand, we consider outward-propagating oscillations at frequency $\omega$:  $R_+^{(0)} = k e^{i\omega t_{i+}}$, where $k$ is a constant amplitude and the real part is implied.  The first-order reflected wave satisfies 
\[
\drop \calR_-^{(1)} =  \calR_+^{(0)}\partial_{r_0} \ln \sqrt{Z}
= k e^{i\omega (t-\tau_0)} \partial_{r_0} \ln \sqrt{Z}, 
\]
and has zero amplitude at $r_0=\infty$.  We use the relation $t = t_{i-} - \tau_0$, appropriate to the inward characteristic, to eliminate $t$:
\[
\drom \calR_-^{(1)}
= k e^{i\omega( t_{i-} - 2\tau_0)} \partial_{r_0} \ln \sqrt{Z}. 
\]
Integrating along the in-going characteristic with the constraint that  $\calR_-^{(1)}=0$ at $r=\infty$, 
\[ 
\calR_-^{(1)}(r_0) = k e^{i\omega t_{i-}} \int_{r_0}^\infty e^{-2i\omega\tau_0(r')} \partial_{r_0} \ln \sqrt{Z(r_0')}\, dr_0'; 
\] 
and changing integration variables from $r_0$ to $\tau_0$ using $dr_0=\cso d\tau_0$, 
\begin{equation}\label{eq:Highfreq-reflection} 
\calR_-^{(1)}(\tau_0) = k e^{i\omega t_{i-}} \int_{\tau_0}^\infty e^{-2i\omega\tau_0'}  \partial_{\tau_0'} \ln \sqrt{Z(\tau_0')}\, d\tau_0'. 
\end{equation}

In this form, we recognize that the reflected amplitude is related to the Fourier transform of $\partial_{\tau_0} \ln \sqrt{Z(\tau_0)}$ evaluated at frequency $2\omega$.  The adiabatic invariance of $\calR_\pm$   at high frequencies, therefore, arises from the fact that this Fourier amplitude is very small when the wave frequency is much higher than the characteristic frequencies of the reflecting structure.   Although equation (\ref{eq:Highfreq-reflection}) is only the first step in an iterative solution to equation (\ref{eq:acoustics-calR}), it is increasingly accurate in the high-frequency limit because of the diminishing amplitude of the reflected wave.

\section{Nonlinear acoustics} \label{S:NonlinearAcoustics}

Our applications of interest involve the nonlinear effects of wave self-distortion, shock formation, and shock evolution. So, we now consider wave motion in the nonlinear regime.  We begin by demonstrating the existence of nonlinear solutions to certain problems in the high-frequency limit.  We then evaluate the nonlinear error associated with Lighthill's proposal that the linear invariant may be used to approximate nonlinear problems.  Finally, we consider the validity of Lighthill's shock formation criterion. 

\subsection{Nonlinear solutions}\label{SS:NonlinearSolutions}
 
In certain cases, we can find nonlinear solutions.   The nonlinear equation (eq.~\ref{eq:dJdr_Unlinearized}) is integrable provided, first, that reflections are negligible so a traveling wave can self-consistently be considered quasi-simple ($\delta J_\mp=0$, so $v=\pm \delta c/q$), either by virtue of the high-frequency limit or because $Z$ is constant; and second, that the  environmental variables ($A,\rho_0,\cso$) obey some power law relationship while $q$ is constant. 

For a quasi-simple wave, equation (\ref{eq:dJdr_Unlinearized}) becomes 
\begin{equation}\label{eq:dJdrNonlnSimple} 
\dropm {\delta j}_\pm = w_{A\pm} \partial_{r_0} \ln A + w_{\rho\pm} \partial_{r_0} \ln \rho_0 + w_{c\pm} \partial_{r_0} \ln c_0 + w_{q\pm} \partial_{r_0} \ln q, 
\end{equation} 
 where ${\delta j}_\pm = \delta J_\pm/c_0$ is the normalized perturbation, 
 \begin{eqnarray}\label{eq:defwA}
 w_{A\pm} &=& - \frac{{\delta j}_\pm}{2} f_\pm^{-1/q},\\
 \label{eq:defwrho}
 w_{\rho\pm} &=& \pm \frac{1}{\gamma} \left(f_\pm^{-\frac{1+q}{q}} - f_\pm\right), \\ 
 \label{eq:defc}
 w_{c\pm} &=&   2 w_{\rho\pm} - \frac{{\delta j}_\pm}{2}, 
 \end{eqnarray} 
 and 
 \begin{equation}
  w_{q\pm} = -\frac{{\delta j}_\pm}{2} \left(1-\frac{2q^2}{\gamma^2}\right) 
  \pm \frac{2q}{\gamma^2} \left(1-f_\pm^{-\frac{1+q}{q}}\right) 
  \pm  \left(\frac{f_\pm}{q} - \frac{2}{\gamma} f_\pm^{-\frac{1+q}{q}} \right) \ln f_\pm,
 \end{equation}
where we define 
 \begin{equation} 
 f_\pm = 1 \pm \frac{q}{2} {\delta j}_\pm. 
\end{equation}  

Adding the requirements that $q$ is constant, and that there exist power law relationships between $A, \rho_0$, and $c_0$ so that their logarithmic derivatives are linearly related, renders the problem integrable.   Let $X(r_0)$ be any function of position, and let $A(r_0)\propto X(r_0)^{k_A}$, $\rho_0(r_0)\propto X(r_0)^{k_\rho}$, and $c_0(r_0) \propto X(r_0)^{k_c}$ for  constants $k_A$, $k_\rho$, and $k_c$.  Then, equation (\ref{eq:dJdrNonlnSimple}) is solved when the quantity
\begin{equation}\label{eq:NonlinSoln} 
    X(r_0)~ \exp\left[ \int
    ^{{\delta j}_\pm} 
    \frac{d\, {\delta j}}{k_A w_{A\pm} + k_\rho w_{\rho\pm} + k_c w_{c\pm}}
    \right]
\end{equation}
is conserved along the active  characteristics -- that is, along $C^+$  if the wave is traveling outward, or along $C^-$ if it is inward.  
(Note, the integral diverges as its arbitrary lower bound tends to zero, although expression (\ref{eq:NonlinSoln}) converges in this limit.) 

For a concrete example, consider the case of a uniform fluid in which only $A$ varies along the channel.  This corresponds to $X(r_0)=\sqrt{A(r_0)}$, $k_A=2$, and $k_\rho=k_c=0$ above.  Using equation (\ref{eq:NonlinSoln}) and the above definitions, we see that 
\[
    \sqrt{A}~ \exp\left[ \int^{\delta J_\pm}  \left(1\pm\frac{q}{2} \frac{\delta J}{\cso}\right)^{1/q} \frac{d\,\delta J}{\delta J} \right]
    \]
is conserved along $C^\pm$ in this case.   
When the nonlinear term is negligible, this implies that $\calR_\pm$ is conserved along $C^\pm$ in the high-frequency limit.  The novel element is a nonlinear correction to the effective wave amplitude, which also modifies the propagation speed. 

The conservation of $\calR_\pm$ for small-amplitude perturbations in this example is not surprising, given our linear results in \S\,\ref{SS:calRandLw}.   Indeed, one can reconstruct this linear result from equation  (\ref{eq:NonlinSoln}) by noting that $w_{A\pm}=w_{\rho\pm} = -{\delta j}_\pm/2$ and $w_{c\pm} = -3{\delta j}_\pm/2$ to first order in ${\delta j}_\pm$, while $w_{q\pm}=0$ to the same order.   

A counterexample is the special case in which $A \rho_0 c_0^3$ is independent of $r_0$ (i.e., $k_A+k_\rho+3k_c=0$), for which the conservation of $\calR_\pm$ would imply  that ${\delta j}_\pm$ is constant.  In this case, the denominator in equation (\ref{eq:NonlinSoln}) is zero to linear order in ${\delta j}_\pm$; the nonlinear terms then imply that  ${\delta j}_\pm$ varies logarithmically with $X(r_0)$ even for small perturbations.    Because the conservation of $\calR_\pm$ can be understood in terms of energy conservation, and because our numerical experiments for this case are consistent with the hypothesis that  $\calR_\pm$ is conserved for small perturbations, we defer a full analysis to later work.   

\subsection{Nonlinear error incurred by fixing the linear adiabatic invariant} \label{SS:NonlinError}
The essence of Lighthill's proposal is that an adiabatic invariant derived from linear acoustics, such as $\calR_\pm$, remains a useful approximate invariant for nonlinear acoustics.  It can therefore be used to evaluate wave self-distortion, shock formation, and weak shock evolution.  

We have already seen that the nonlinear invariant of equation (\ref{eq:NonlinSoln}), when it exists, reduces to $\calR_\pm$ when nonlinear terms are negligible (at least, when $A \rho_0 c_0^3$ varies with $r_0$).  To address the more general case in which we cannot obtain a nonlinear solution,  we compare the nonlinear equation for $\dropm J_\pm$, equation (\ref{eq:dJdr_Unlinearized}), with what one would derive from equation  (\ref{eq:acoustics-calR}).  To leading order in $v/\cso$ and $\delta x$, the residual is 
\begin{equation}\label{eq:NonlinearResidual}
v\,\delta x \,\partial_{r_0}\ln A \pm \cso (1+q)(\delta x)^2 \partial_{r_0}\ln \left[\cso (1+q)^{3/2} \rho_0^{1/2}\right],
\end{equation}
which for quasi-simple waves ($\delta v= \pm \cso\, \delta x$ to leading order)  further simplifies to 
\begin{equation} \label{eq:NonlinErrorSimple}
\pm \cso (1+q)(\delta x)^2 \partial_{r_0} \ln \left[A^{1/(q+1)} \cso (1+q)^{3/2} \rho_0^{1/2} \right]. 
\end{equation} 
In addition to being second-order in $\delta x$ and $v/\cso$, this  is proportional to logarithmic gradients of the background quantities.  From this we infer that, if the background quantities vary on a scale $H$, a waveform with peak velocity $v_1$ that is predicted from equation (\ref{eq:acoustics-calR}) will be spoiled by nonlinear distortions after it has propagated a distance $\sim \cso H/v_1$.  However, shock formation occurs after only $\sim \cso/v_1$ wavelengths ($\cso\lambda/v_1$, using `$\lambda$' here to mean wavelength rather than wave speed), so the relative nonlinear error at the wave peak is only of order $\lambda/H$ at shock formation.  

\subsection{Shock formation}\label{SS:ShockFormation}
We turn to the criterion for shock formation.  Importantly, shocks form within waves at local maxima of the compression rate -- at or near a nodes of $v$, rather than its peaks.   However,  nonlinear distortion, which is of order $v/\cso$, is zero at the node.  Lighthill's criterion for shock formation is derived by assuming that the invariant from linear theory (e.g., our $R^\pm$) is conserved along $C^\pm$ characteristics.  So long as wave reflection is also negligible and the medium is otherwise still, so that the conditions for a quasi-simple wave are met, Lighthill's criterion for shock formation is exact.  

For example, there is no reflected amplitude at the leading edge of a wave travelling into a still medium so long as $Z$ is continuous.   Lighthill's criterion  is, therefore, exact for the formation of a `head' shock forming at the wave's leading edge.  To reinforce this point, we demonstrate in the Appendix that Lighthill's proposal reproduces the differential equation obtained in the `wavefront expansion' technique \citep{whitham1975linear} for identifying the shock criterion at a head shock.  As reflection is negligible in the high-frequency limit, Lighthill's criterion is also essentially exact for internal and `tail' shocks within high-frequency waves. 

Even if the background is acoustically active, the only effect at a node of $J_\pm$ is to change the wave speed by $(1-q)J_\mp/2$.  As the average of this quantity is close to zero, it has little effect on the $C^\pm$ trajectories whose crossing indicates shock formation. 

On the other hand, significant wave reflection reduces the strength of gradients within the wave.  Internal and `tail' shocks can, therefore, be delayed or eliminated by sufficiently strong reflection.  Of course, the reflected wave may also create its own shock.

\section{Numerical Tests }\label{S:numerics}

\begin{figure}
  \includegraphics[width=\textwidth]{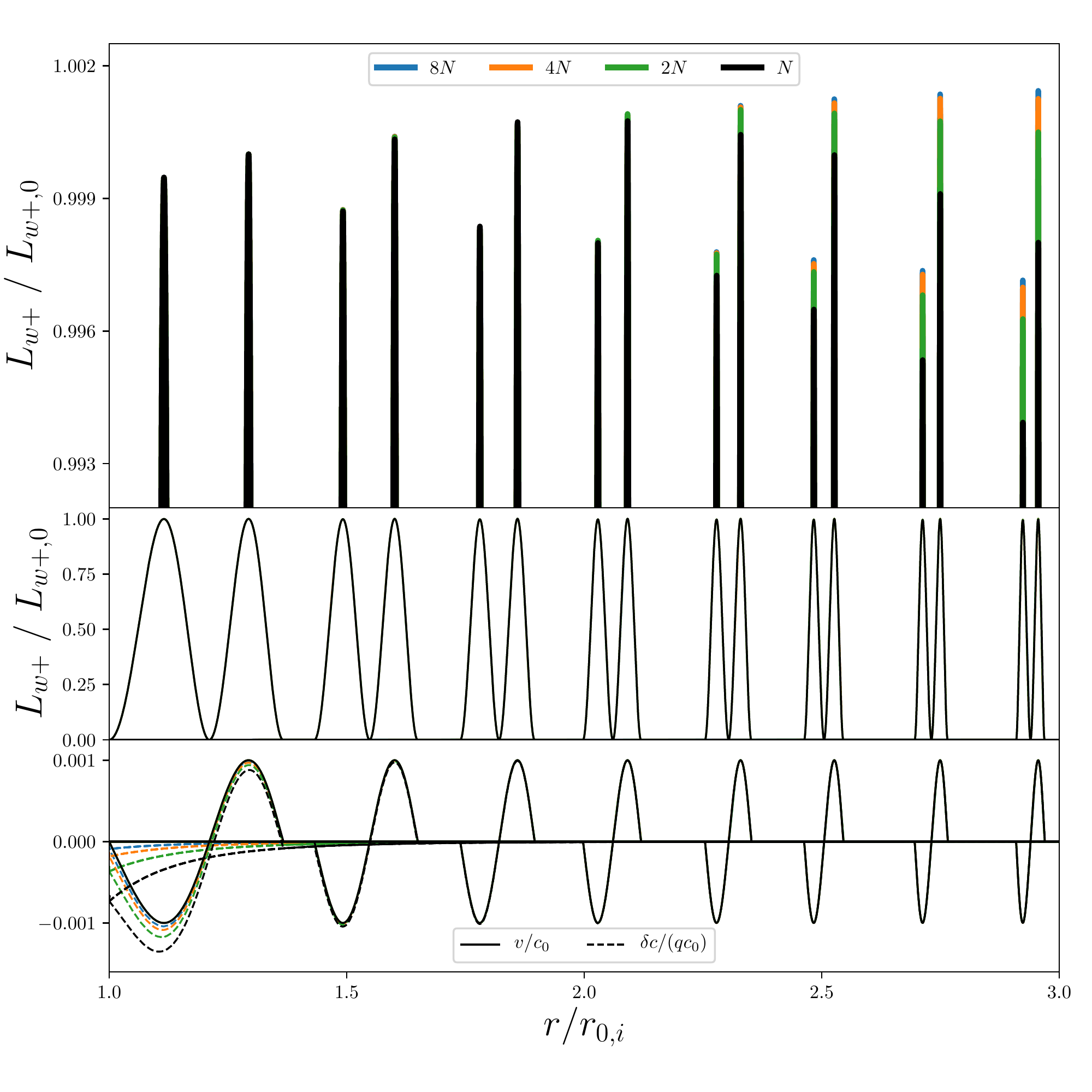}
  \caption{ Wave pattern (bottom panel) and normalized instantaneous wave luminosity (top panels) of a single sinusoidal transient wave propagating in a spherical ($A\propto r^2$), constant-pressure background in which density increases as  $\rho_0(r)\propto r_0^4$ such that the impedance is constant.  Multiple snapshots are over-plotted to indicate the transient's progression.  Each coloured line represents a different simulation resolution, where the number of uniformly-spaced grid cells is listed in units of the base resolution $N=8192$. The velocity perturbation is small, less than $10^{-3}\cso$ over the domain. The top panel provides an expanded view of the middle panel to show how the small deviations from perfect wave luminosity conservation depend on numerical resolution. }
\label{fig:flashNoRefl}
\end{figure}

We  present three numerical tests covering the phases before, at, and after shock formation.  We use the hydrodynamics code \texttt{FLASH4.6} \citep{flashcode} set to use the HLLC Riemann solver with fifth-order reconstruction.  The spherical simulation domain ($A = 4\pi r^2$) spans $1\leq r/r_{0,i}\leq 3$ with uniform radial resolution. The initial fluid pressure is constant, but the density increases as $\rho_0\propto  r^4$ such that $Z$ is constant. 
We adopt an ideal gas equation of state with $\gamma=5/3$.  Variations in pressure, density, and velocity related by $v=\delta c / q$ are applied at the inner boundary to generate an outward traveling wave.

First, we test the prediction that there is no wave reflection in a region of uniform impedance, so wave luminosity should be conserved along characteristics.  Here, we introduce one period of a sinusoidal wave (which starts in compression), and set its velocity amplitude to $10^{-3}c_0$ at the inner boundary.  We choose $\omega$ so the acoustic wavelength is twice the density scale height (twice $r/4$) at the inner boundary;  however, the drop in $c_0$ causes the wavelength to be significantly smaller than $r/4$ at the outer boundary. 
Figure \ref{fig:flashNoRefl} shows the normalized instantaneous wave luminosity $L_{w+}$ from four simulations with resolution that increases relative to a base resolution of $N=8192$ cells.  Wave maxima and minima both display peaks of $L_{w+}$, which are slightly higher at the maxima. We see that the deviations from constant luminosity in the wave maxima (minima) decreases with increasing spatial resolution to a maximum of $0.02\%$ ($0.4\%$) for the highest resolution simulation. The numerical results support the conclusion that luminosity is conserved along characteristics,  across nearly two orders of magnitude variations in background density, for a wave traveling in a fluid with constant impedance. 

\begin{figure}
  \includegraphics[width=\textwidth]{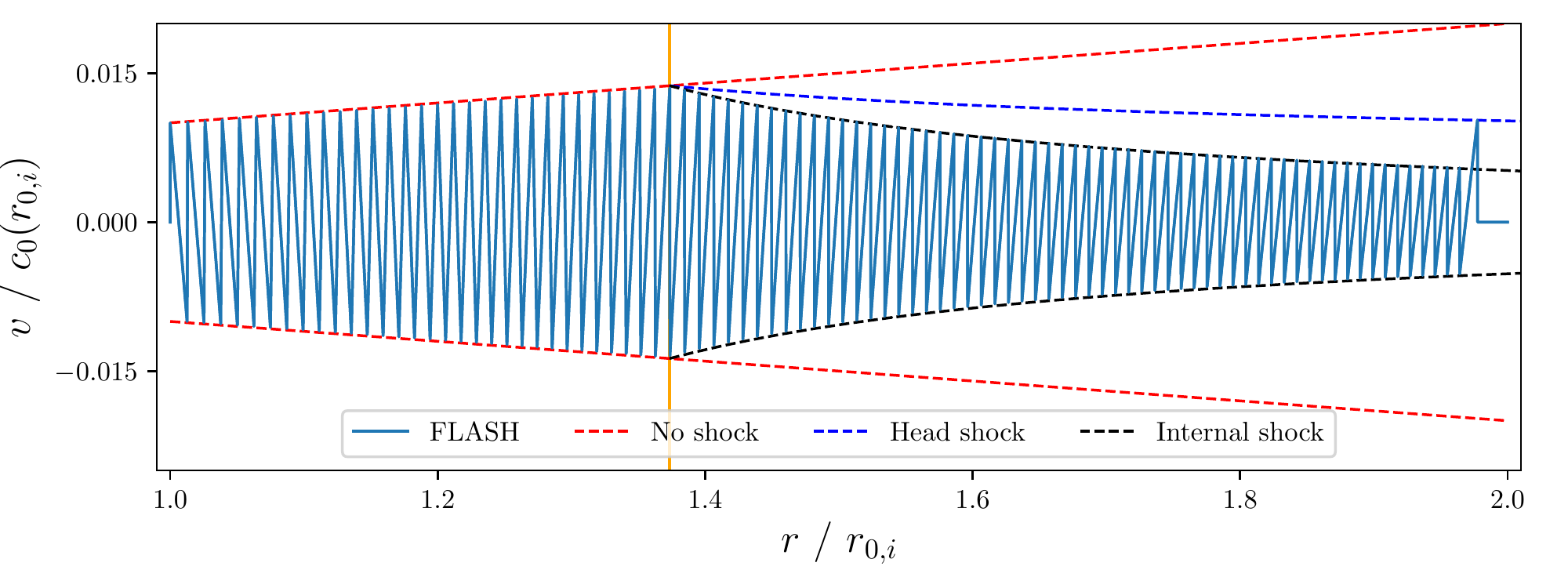}
  \caption{\cdmnew One snapshot of shock evolution within a wave with an initial `reverse saw-tooth' waveform.  The simulation matches analytical predictions for the shock formation radius (orange vertical line) and for the velocity amplitudes of the head shock and internal shocks (blue and black dashed lines, respectively) to high accuracy.  These predictions, discussed in depth in a companion paper \citep{MatznerRo20ApJ}, rely on the adiabatic invariance of $\calR_+$. } 
\label{fig:flashShockFormn}
\end{figure}

{\cdmnew 
Next, we check Lighthill's proposition that mildly nonlinear wave distortion can be accurately predicted using the linear adiabatic invariant.   Using the same background profile and initialization strategy, and adopting the highest resolution, we increase the amplitude of the sinusoidal wave to $10^{-2}c_0$ while doubling the wave frequency, so that shock formation occurs within the domain.  In Figure \ref{fig:flashShockFormn}, we compare numerical and analytical predictions for the form of the wave pulse at the time of shock formation.  The two agree very well, at least at the level estimated in \S~\ref{SS:NonlinError}, despite the fact that the wave has traversed over two orders of magnitude in background density from the point of initialization. }

\begin{figure}
  \includegraphics[width=\textwidth]{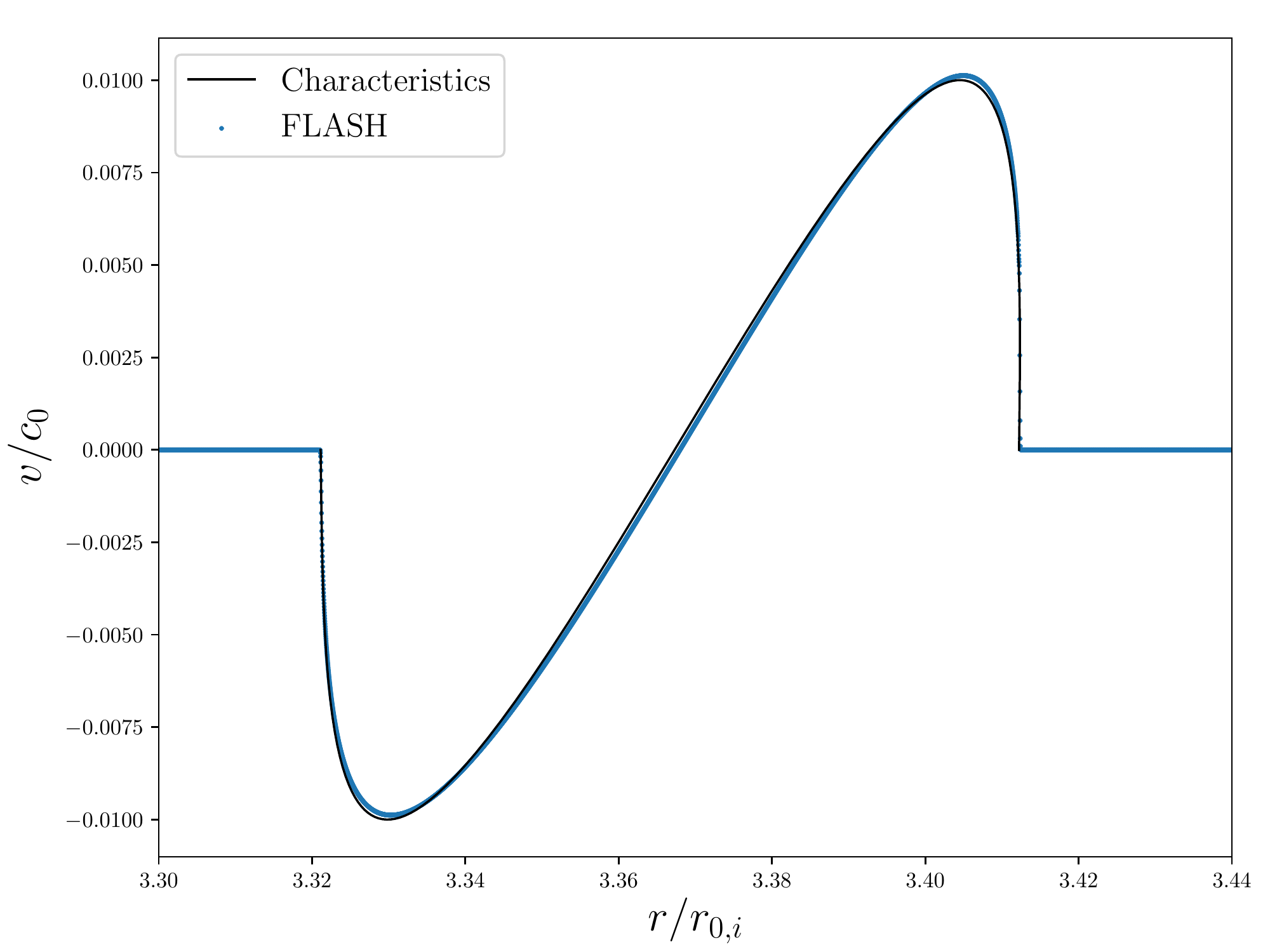}
  \caption{\cdmnew Analytical estimate (blue curve, obtained by conserving $\calR_+$ along $C^+$ characteristics) and numerical result (black dots) for a wave at the point of shock formation.  The wave amplitude is initialized to $10^{-2}c_0$ and the wavelength is initially $r/2$. Other aspects of the simulation match those in the highest-resolution runs shown in Figure \ref{fig:flashNoRefl}.  From initialization to shock formation, the wave has traversed a range of densities varying by a factor of $3.41^4=135$. }
\label{fig:SawtoothWave}
\end{figure}

{\cdmnew
For a final demonstration, we show in Figure \ref{fig:SawtoothWave} that the adiabatic invariance of  $\calR_\pm$ can be used to adapt \citeauthor{whitham1975linear}'s equal-area method to weak shock evolution within a varying background.  Here, a wave train is introduced for which the initial waveform is a `reverse saw-tooth' with a linear rise and sudden decline in $\calR_+(t_0)$.  The waveform converts to a `forward saw-tooth' at the point of shock formation.   After shock formation, the wave dissipates in a manner we discuss in detail in a companion paper \citep{MatznerRo20ApJ}.  Notably, the leading or `head' shock decays distinctly more slowly than the `internal' shocks that form within the wave train. 
}

\section{Summary and conclusions}

Our primary findings are as follows:
\begin{itemize}
    \item[-] The linearized equations of hydrodynamics are compactly expressed (eq.~\ref{eq:acoustics-calR}) in terms of wave amplitudes $\calR_\pm$, which 
    interact in the presence of variations of the impedance. 
    \item[-] In the context of `quasi-simple' waves (\S~\ref{SS:QuasiSimple}) -- those in which one wave family is absent and not regenerated by reflection -- the conservation of $\calR_\pm$ is consistent with the invariant proposed by \citet{lighthill2001WavesInFluids}, and also with the conservation of wave luminosity (\S~\ref{SS:Lw}).  
    \item[-] Familiar results about the dynamics of reflection follow directly from the amplitude equation (\S\ref{S:Reflection}).  In particular, wave reflection is strongly suppressed in the high-frequency limit.  
    \item[-]  Lighthill's proposition amounts to the statement that linear conservation laws may be extrapolated into the mildly nonlinear regime, providing a simple means to calculate wave self-distortion and shock formation, as well as weak shock dissipation by Whitham's equal-area method.  We validate this method and delimit its validity. In the high-frequency limit, the nonlinear error is minimal at the point of shock formation (\S~\ref{SS:NonlinError}).  Because there is no nonlinear error at the wave node, Lighthill's  criterion for shock formation is exact so long as the shock forms at a node and reflection is negligible, as it is for `head' shocks, and for  any shocks generated within high-frequency waves (\S~\ref{SS:ShockFormation}). 
    \item[-] Going beyond linearized theory, we find that nonlinear solutions are available in a restricted class of problems (\S~\ref{SS:NonlinearSolutions}).  In the example considered, the novel element is a nonlinear correction to the wave amplitude, which leads to a modified phase speed. 
\end{itemize}

In a companion work \citep{MatznerRo20ApJ}, we use these findings to predict the evolution of shock strength and shock dissipation in the context of super-Eddington outbursts of evolved massive stars.  There, we provide useful formulae for the shock-deposited heat, and we evaluate the potential of head shocks to eject matter from the stellar surface. 

\vspace{0.3cm}

 This work was funded by an NSERC Discovery Grant (CDM) and by the  Gordon and Betty Moore Foundation through Grant GBMF5076 (SR).  

\vspace{0.1cm}
\noindent{\em Software:}  \texttt{FLASH4.6} \citep{flashcode}.

\vspace{0.1cm}
\noindent{\em Greenhouse gas budget:} We estimate 50 kg CO$_2$ equivalent arising from the presented simulations, based on $\sim$200\,kWh of electricity in Berkeley, CA. 

\vspace{0.1cm}
\noindent Declaration of Interests: the authors report no conflict of interest.

\appendix
\section{}\label{Appendix}

{\bfa 
We wish to demonstrate the  differential equation involved in identifying shock formation by the `wavefront expansion method' \citep{whitham1975linear} is reproduced by Lighthill's proposition -- i.e., by assuming that $\calR_+$, or equivalently $L_{w+}$, is conserved  along $C^+$ characteristics, and that the equipartition condition holds.  

The velocity downstream of an outward-traveling wavefront located at $r_{0+}(t)$ is, to first order, 
\begin{equation}\label{eq:appendix_u}
v = v_0 + \xi v_1,
\end{equation}
where $\xi=r-r_{0+}(t)$ is the distance behind the wavefront, $v_1=\pt_r v$ is the partial spatial derivative of $v$, and $v_0=0$ for a static background. (Here, $r$ and $r_{0+}$ do not refer to the same fluid element.) Note that $\pt_t \xi = -c_0$ and $\pt_r \xi = 1$. 

We may neglect reflection in the vicinity of the wavefront as the upstream fluid is considered to be undisturbed in the wavefront expansion technique (see \S~\ref{SS:ShockFormation}).  Given that  counter-propagating waves are negligible,  equipartition  holds ($\delta c = q v$) and characteristics travel at the speed
\begin{equation}
    v+c = c_0 + \frac{\gamma+1}{2}\xi v_1. \label{eq:appendix_u+c}
\end{equation}
From equations (\ref{eq:def-Lw}) and (\ref{eq:appendix_u}), the wave luminosity of a characteristic is 
\begin{equation}\label{eq:appendix_lwave}
    L_{w+}= \xi^2 A \rho_0 c_0  v_1^2,
\end{equation}
with partial derivatives
\begin{eqnarray}
\pt_t L_{w+} &=& \xi A\rho_0c_0v_1\left(-2c_0 v_1 + 2 \xi v_1' \right), \label{eq:appendix_ptt_lwave} \\
\pt_r L_{w+} &=&  \xi A \rho_0 c_0 v_1\left(\xi v_1 \pt_r{\rm ln}\left( A \rho_0 c_0 \right) + 2 v_1 \right)\label{eq:appendix_ptr_lwave}.
\end{eqnarray}
Substituting (\ref{eq:appendix_u+c})-(\ref{eq:appendix_ptr_lwave}) into (\ref{eq:acoustics-Lw}) and retaining lowest-order terms (acceptable because of a vanishing nonlinear error; see  \S~\ref{SS:NonlinError}) yields the Ricatti equation for $v_1$
\begin{equation}
    0 = v_1 + \frac{c_0}{2} \pt_r{\rm ln}\left( A \rho_0 c_0 \right)  v_1 + \frac{\gamma+1}{2}v_1^2
\end{equation}
previously obtained by \citet{tyagi2005propagation} and \citet{2017ApJ...841....9R} using the wavefront expansion technique. 
}

\bibliographystyle{jfm}
\bibliography{LeadingWaveRefs}

\end{document}